\documentclass[preprint]{JHEP3} 

\usepackage{epsfig,multicol,youngtab}

\newcommand\fverb{\setbox\pippobox=\hbox\bgroup\verb}
\newcommand\fverbdo{\egroup\medskip\noindent%
			\fbox{\unhbox\pippobox}\ }
\newcommand\fverbit{\egroup\item[\fbox{\unhbox\pippobox}]}
\newbox\pippobox

\newcommand{\be}{\begin{equation}} 
\newcommand{\ee}{\end{equation}}

\newcommand{\ba}{\begin{eqnarray}}
\newcommand{\ea}{\end{eqnarray}}

\title{Strong coupling anomalous dimensions of ${\cal N} = 4$ super Yang-Mills}

\author{Matteo Beccaria\thanks{Partially supported by INFN, IS-RM62}\\
	Dipartimento di Fisica, Universita' di Lecce,\\
        Via Arnesano, 73100 Lecce\\
	INFN, Sezione di Lecce\\
	E-mail: \email{matteo.beccaria@le.infn.it}}

\author{Carmine Ortix\thanks{Partially supported by INFN, IS-RM62}\\
	Dipartimento di Fisica, Universita' di Lecce,\\
        Via Arnesano, 73100 Lecce\\
	INFN, Sezione di Lecce\\
	E-mail: \email{carmine.ortix@le.infn.it}}

\abstract{
We study the strong coupling behaviour of fixed length single trace operators in the scalar SU(2) sector of ${\cal N} = 4$ SYM.
We assume the recently proposed connection with a twisted half-filled Hubbard model.
By explicit direct diagonalization of operators with length $L=4, 6, 8$ we study the full {\em perturbative multiplet} of 
those lattice states which have a clear correspondence with gauge theory composite operators.
For this multiplet, we follow the weak-strong coupling flow to free fermion states and identify in particular the 
precise asymptotic fermion configuration.
Next, we analyze the Lieb-Wu equations of the twisted Hubbard model. For the antiferromagnetic state we derive its strong coupling expansion 
working at $L$ up to $32$. We also study the lightest state in the perturbative multiplet. This state is non trivial
since involves complex solutions of the Lieb-Wu equations. It is particularly interesting for $AdS_5\times S^5$ duality 
since it is dual to the folded string semiclassical solution in the thermodynamical limit. 
We are able to perform the full analysis and compute the next-to-next-to leading terms in the strong coupling expansion 
for the non trivial lengths $L=12$ and $L=20$. A general formula is proposed for the NLO expansion 
for any $L=4(2k+1)$, $k\in\mathbb{N}$.
}


\begin{document} 

\section{Introduction}
\label{Sec:Intro}

The quantum behavior of ${\cal N} = 4$ super Yang-Mills in the planar limit is crucial in the context of 
AdS/CFT correspondence~\cite{Maldacena:1997re,Gubser:1998bc,Witten:1998qj,Kazakov:2004qf,Klebanov:2000me}.
In particular, the anomalous dimensions of certain single trace operators 
in the planar limit of the ${\cal N} = 4$ theory can be compared with the masses of string states on $AdS_5\times S^5$~\cite{Plefka:2005bk}. 
For instance, the comparison turns out to be particularly favorable for BMN states~\cite{Berenstein:2002jq} where the gauge-string matching
can be done at the perturbative level in the planar gauge theory at the price of analyzing {\em long} composite operators with a large number of 
constituent fields and few impurities.

Apart from the gauge-string connections, the ${\cal N} = 4$ quantum theory has a rich internal structure suggesting its 
quantum integrability. The calculation of anomalous dimensions in specific sectors of the ${\cal N} = 4$ theory can be cast
in algebraic form by computing the loop corrected dilatation operator~\cite{Beisert:2003tq}. The huge mixing problem is then reduced
to the analysis of the eigensystem of the finite dimensional matrix representing the dilatation operator.

Remarkably, at one-loop, the dilatation operator can be identified with the Hamiltonian of the integrable XXX spin $1/2$ lattice model~\cite{Minahan:2002ve}. 
In the SU(2) sector, Beisert, Dipple and Staudaucher (BDS)~\cite{Beisert:2003tq,Serban:2004jf,Beisert:2004hm}
proposed a Bethe Ansatz for the 2-body $S$-matrix in agreement with the explicit three loop dilatation operator and consistent with 
all loop BMN scaling~\cite{Berenstein:2002jq}. The BDS equations describe a spin model of the Heisenberg-type with long range couplings.
The range of the spin interaction increases as the loop order is increased. After having built the five loop BDS Hamiltonian, they 
could match the gauge theory predictions known up to three loops. A disagreement with the gauge theory calculation for operators with classical 
dimension $L$ is expected precisely at $L$ loop order due to wrapping terms. In the thermodynamical limit, these terms are negligible and 
an all-loop Bethe Ansatz was proposed.

Using the BDS equations it is possible to compute the largest energy state on the chain~\cite{Rej:2005qt,Zarembo:2005ur}. 
In the Heisenberg model language, it the non trivial antiferromagnetic state. Its energy in the thermodynamical limit perfectly agrees with 
the ground state energy of a (twisted)  one-dimensional  Hubbard model at half-filling~\cite{LiebWu}. 
The connection between the spin model and the itinerant fermion model can be understood as follows. 
The spin model is nothing but the strong coupling expansion of the fermion model, where strong means the the hopping term is 
treated perturbatively. The effective Hamiltonian for the strong coupling of the Hubbard model contains interactions with a range increasing with the 
order of the expansion. This is because successive applications of the hopping operators connect lattice sites with increasing distance.

With an impressive breakthrough, Rej, Serban and Staudacher (RSS) proposed the Hubbard model as the correct {\em microscopic}  model behind the integrable structure of the 
${\cal N} = 4$ SYM dilatation operator~\cite{Rej:2005qt}. In other words, they suggest that it could predict at all loops and non perturbatively the anomalous dimensions of the 
gauge theory operators for any finite $L$. This proposal also overcomes the problems related to the wrapping interactions in the long range spin model~\cite{Ambjorn:2005wa}.
The RSS proposal is still a conjecture although with robust theoretical motivations. To pursue its assessment, it would be necessary to perform
a four loop calculation in the gauge theory. Waiting for this check, the Hubbard model can be considered as a powerful Ansatz for the description 
of the gauge theory at finite operator lengths and beyond the perturbative regime.

Actually, the appearance of the Hubbard model remains somewhat mysterious and intriguing~\cite{Mann:2006it}. In particular, the Hubbard model describes
fermions with spin and admits states with two fermions in the same site. This extra states cannot be identified immediately with gauge theory
operators. Indeed, it is not clear what is the role of its extra states which at strong coupling definitely mix with the perturbative states.
Waiting for a better understanding of the role of Hubbard model, we can assume an optimistic attitude and exploit it 
to investigate the weak-strong coupling behavior of anomalous dimensions. 

The main technical tool in the analysis of states in the Hubbard model are its Lieb-Wu equations which encode integrability.
The Lieb-Wu equations are precious in the study of the thermodynamical limit including finite size corrections~\cite{Hernandez:2005nf,Beisert:2005mq,Feverati:2006tg}. 
However, the machinery works well only for particular states. In the thermodynamical limit, the Bethe roots accumulate on a discrete set of non trivial curves
in the complex plane. The integral equations for the root density are quite difficult due to unknown shape (and number) of the contours.
At weak coupling, the Lieb-Wu equations reduce to the BDS Ansatz of Heisenberg type. In some favorable cases a solution for the thermodynamical limit can be found. 
Remarkably, the solution can also be matched with specific semiclassical string states~\cite{Beisert:2003xu,Beisert:2003ea,Beisert:2005di,Frolov:2003qc,Park:2005kt}. 
The analysis of the Lieb-Wu equations is more problematic and the full spectrum of the twisted Hubbard model at finite large $L$ remains a quite difficult task.

As an alternative approach to solving the Lieb-Wu equations, J. Minahan has recently proposed~\cite{Minahan:2006zi} to analyze directly the twisted Hubbard model 
Hamiltonian on small lattices to understand the features of the spectrum. This is an interesting approach aimed at understanding the strong coupling 
behavior of states associated to gauge invariant operators. 

In this paper, we analyze single trace cyclic operators with zero SU(2) spin of the form 
\be
\mbox{Tr}\left(Z^{J}\,\Phi^{J}\right) + \dots
\ee
with large $J$. Here, $Z$, and $\Phi$ are the charged scalar fields in the ${\cal N} = 4$ supermultiplet. 
As usual, dots stand for various other orderings of the scalar fields required to obtain eigenstates of the dilatation operator. 
The above set of operators, all with classical dimension $2J$ is closed under perturbative renormalization. It is not clear what
happens at strong coupling. Indeed, it has been suggested that some surprise could occur~\cite{Minahan:2005jq}.
In this direction, it seems to be important to understand the mutual role of the perturbative states and the additional states in the Hubbard model. 

As a first step of our analysis, we collect and explain some general features of the states belonging to what we call the {\em perturbative multiplet}.
These are
the states in the Hubbard model with a clear identification with single trace operators in the gauge theory. Then, we exploit the direct 
approach of~\cite{Minahan:2006zi} to study the full spectrum at $L=4, 6, 8$. The case $L=4$ is discussed in~\cite{Minahan:2005jq}
and here it is reviewed to present the method and prepare for larger $L$. The cases $L=6$ and $8$ are more involved and reveal
interesting features. In all cases, we determine precisely the flow to large $g$.

The direct analysis can hardly be pushed to much larger values of $L$. Hence, we revert to the numerical exploration of the 
complete Lieb-Wu equations. We perform the analysis of the antiferromagnetic state
(the one with highest anomalous dimension) working with up to $L=32$ sites and providing the NNLO strong coupling expansion of its
anomalous dimension. 

Next, we consider the lightest state in the perturbative multiplet. Exploiting some features of this state in the thermodynamical
limit, we solve numerically the Lieb-Wu equations at $L=12, \,20$ (the first non trivial cases of $L=8k+4, k\in\mathbb{N}$). 
These are remarkable values of the composite operators length. Nevertheless, the analysis is performed in full details
obtaining also in this case the NNLO strong coupling expansion of the anomalous dimension. This analysis is important in our opinion
since it is an explicit study of the Lieb-Wu equations at large but finite $L$.

The paper is closed by a conjecture about the general form of the lightest state for lengths $L$ of the form $4(2k+1)$, $k\in\mathbb{N}$.
A discussion of the comparison with the BMN limit is also discussed.

\section{Anomalous dimensions in ${\cal N}=4$ SYM and the Hubbard model}

The four dimensional ${\cal N}=4$ SYM theory is finite. Its non trivial quantum properties
are encoded in the behavior of gauge invariant composite operators. The gauge group is $SU(N)$ and 
we are interested in the planar limit $N\to \infty$. We introduce
the coupling $g$ related to the large $N$ 't Hooft coupling $\lambda$
\be
g^2 = \frac{\lambda}{8\pi^2},\qquad \lambda = g^2_{YM}\,N.
\ee
In the so-called SU(2) sector of ${\cal N}=4$ SYM, we consider gauge invariant
composite operators of definite scaling dimension 
\be
\mbox{Tr}\left(Z^{L-M}\,\Phi^{M}\right) + \dots
\ee
where $Z$, $\Phi$ are charged scalar fields. The classical dimension is $L$, the number of fields. In the following, we shall simply call $L$
the {\em length} due to the lattice representation that we are going to introduce. The above composite operators
have non trivial renormalization properties and acquire anomalous dimensions with all-loop corrections
\be
\Delta(g) = L + \sum_{\ell\ge 1} \Delta_\ell\, g^{2\ell}.
\ee
As we discussed in the Introduction, they can be computed as the 
eigenvalues of a charge ${\cal D}$. It is the dilatation operator and belongs to an infinite tower of commuting charges. 
Its perturbative expansion (at two loops) is 
\be
{\cal D} = L + \frac{g^2}{2}\sum_i(1-\sigma_i\cdot\sigma_{i+1})-\frac{g^4}{4}\sum_i(3-4\sigma_i\cdot\sigma_{i+1}+\sigma_i\cdot\sigma_{i+2}) + \cdots,
\ee
where the $\sigma_i$ matrices act on the cyclic states of a spin $1/2$ chain. A particular operator is mapped to a spin chain state
in the natural way, {\em e.g.}
\be
\mbox{Tr}\left(Z\,Z\,\Phi\,Z\,\cdots\right) \longrightarrow |\uparrow\uparrow\downarrow\uparrow\cdots\rangle + \mbox{cyclic translations}
\ee
The dilatation operator at finite $g$ is thus non local. The RSS proposal identifies the up and down spin with fermions in two spin states. It also 
add states with two fermions occupying the same site. On this enlarged state space, RSS define an Hubbard-type Hamiltonian.
In the following we shall consider the case $L\in 2\mathbb{N}$ and $M=L/2$ and look at operators with zero  SU(2) spin.
The explicit local Hubbard Hamiltonian is 
\be
\label{eq:Hamiltonian}
H = H_0+\frac{g}{\sqrt 2}\,H_1,
\ee
where
\ba
H_0 &=& L-\sum_{i=1}^L n_{\uparrow, i}\,n_{\downarrow, i}, \qquad n_{\sigma, i}^{\phantom\dagger} = c^\dagger_{\sigma, i}\,c_{\sigma, i}^{\phantom\dagger},\\
H_1 &=& \sum_{\sigma=\uparrow, \downarrow}\,\left(\sum_{i=1}^{L-1} c^\dagger_{\sigma, i}\,c^{\phantom\dagger}_{\sigma, i+1} + 
e^{i\phi} c^\dagger_{\sigma, L}\,c^{\phantom\dagger}_{\sigma, 1}\right) + \mbox{h.c.}
\ea
The fermion creation and annihilation operators satisfy canonical anticommutation relations
\ba
\{c_{\sigma, i}^{\phantom{\dagger}}, c^{\dagger}_{\sigma', j}\} &=& \delta_{\sigma, \sigma'}\,\delta_{i,j}, \\
\{c_{\sigma, i}, c_{\sigma', j}\} &=& \{c^\dagger_{\sigma, i}, c^\dagger_{\sigma', j}\} = 0.
\ea
Periodic boundary conditions are understood. The twisting phase in the boundary link is fixed at $\phi = \pi/2$.

\noindent
The Hamiltonian (\ref{eq:Hamiltonian}) is symmetric under the SU(2) generators
\ba
S_z &=& \frac{1}{2}\sum_{i=1}^L(n_{\uparrow, i}-\,n_{\downarrow, i}), \\
S^+ &=& \sum_{i=1}^L c^\dagger_{\uparrow, i}\,c^{\phantom\dagger}_{\downarrow, i},\qquad S^- = (S^+)^\dagger.
\ea
At half-filling and with an equal number of up and down fermions the $z$-component is automatically zero.

\medskip
The Hamiltonian (\ref{eq:Hamiltonian}) is also invariant under the shift 
\ba
\label{eq:cyclic}
c_{\sigma, j}&\to& e^{i\phi/L}\ \ \ \ \ \ \,c_{\sigma, j+1},\qquad j=1, \dots, L-1, \\ 
c_{\sigma, L}&\to& e^{i\phi(L+1)/L}\,c_{\sigma, 1},
\ea
with related transformation properties of the $c^\dagger$ operators.
The states invariant under this symmetry will be called {\em cyclic} states and are the
relevant ones to represent single trace operators in the gauge theory.

\medskip
The Hamiltonian can be written in a simpler form after the transformation 
\be
c_n = e^{i\,n\,\phi/L}\,\widetilde{c}_n .
\ee
The on-site Coulombian term $H_0$ is unchanged. The hopping part $H_1$ becomes 
\be
H_1 = \sum_{\sigma=\uparrow, \downarrow}\,\sum_{i=1}^{L} e^{i\phi/L}\,\widetilde{c}^\dagger_{\sigma, i}\,\widetilde{c}^{\phantom\dagger}_{\sigma, i+1} 
 + \mbox{h.c.},
\ee
and the shift symmetry is written simply
\be
\widetilde{c}_n\to e^{2i\phi/L}\,\widetilde{c}_{n+1} .
\ee
The hopping term is diagonalized by introducing fermion operators in  momentum space
\be
a_{\sigma, p} = \frac{1}{\sqrt L}\sum_{n=1}^L e^{-i\,p\,n}\,\widetilde{c}_{\sigma, n},
\ee
where the lattice momenta can take the values
\be
p_n = \frac{2\pi\,n}{L},\qquad n=0, 1, \dots, L-1.
\ee
The dispersion relation for the scaled Hamiltonian $H_1/\sqrt{2}$ is then
\be
\label{eq:dispersion}
\varepsilon_n = \sqrt{2}\,\cos\left(\frac{2\pi n}{L} + \frac{\pi}{2L}\right).
\ee
Cyclic states of the original Hamiltonian with $L$ fermions can be built in momentum spaces by acting on the vacuum with  
$L$ $a^\dagger_{\sigma, p}$ operators with a total momentum $\sum p$ being an odd multiple of $\pi$
due to the phase factor $e^{2i\phi/L}$.

\section{The perturbative multiplet}

As we have seen, the RSS construction introduces additional states with fermion double occupancy which do not have a direct
correspondence with the gauge theory composite operators. As we remarked in the Introduction, the role of these states
is not totally clear. However, we can exploit them as an auxiliary device and focus on what we shall denote
the {\em perturbative multiplet}. This is the set of states which at $g\to 0$ reduce to states with no double
occupancy and have $\Delta(g)\to L$, the maximal value at $g=0$. 
In other words, these are the states which flow at weak coupling to states that can be  naturally mapped to gauge invariant single trace operators.
As $g$  increases, they are no more the maximal energy states and mix with all the extra states. 

At very large $g$, any state flows to a free fermion state which is an eigenstate of the hopping term $H_1$. Our aim 
is to understand the asymptotic free fermion content of specific states in the perturbative multiplet. In particular, at large $g$
we find $\Delta(g)\sim \delta\,\,g + {\cal O}(1)$ where $\delta$ is the hopping energy of the asymptotic free fermion state.
We show in Fig.~(\ref{fig:qualitative}) the qualitative description of the spectrum. The up most state is the so-called
antiferromagnetic state (AF). The bottom part of the perturbative multiplet is composed of what we shall call light states,
where light means that the anomalous dimension is small in the weak coupling region. The multiplets of extra states with 
double occupation are also schematically shown. In principle they can cross the perturbative multiplet.

In the following sections, we shall discuss some general features of the AF and light states that can be derived from 
general principles and an analysis of the BDS equations.

\subsection{The antiferromagnetic state}

At even $L$ the AF state is non degenerate. Its anomalous dimension is known in the thermodynamical limit. In terms
of the planar coupling $\lambda = 8\pi^2\,g^2$, it reads~\cite{Zarembo:2005ur} 
\be
\lim_{L\to\infty}\frac{\Delta_{\rm AF}(\lambda, L)}{L} = 1+\frac{\sqrt{\lambda}}{\pi} f\left(\frac{\pi}{\sqrt\lambda}\right),
\ee
where
\be
f(x) = \int_0^\infty\frac{dk}{k}\frac{J_0(k)\,J_1(k)}{1+e^{2\,k\,x}}.
\ee
This function is well known to be non-analytic in $x=0$. However, it admits the {\em asymptotic} expansion~\cite{Metzner}
\be
f(x) = \frac{1}{\pi}-\frac{x}{4}+\sum_{m = 1}^N \mu_m x^{2m} + {\cal O}(x^{2N+2}),
\ee
where
\be
\mu_m = \frac{(2m-1)(2^{2m+1}-1)\left[(2m-3)!!\right]^3}{2^{3m-1}(m-1)!}\,\frac{\zeta(2m+1)}{\pi^{2m+1}},\qquad (-1)!!\equiv 1.
\ee
Despite being only asymptotic and not convergent, the above expansion has been shown to reproduce correctly 
the second order perturbative correction at large $\lambda$. This means that we can expand the anomalous dimension 
at large $\lambda$ as 
\be
\label{eq:strongaf}
\Delta_{\rm AF}(\lambda, L) = a_1(L)\,\sqrt{\lambda}+a_2(L)+a_3(L)\frac{1}{\sqrt{\lambda}}+\dots,
\ee
and the limits $\lim_{L\to\infty} a_k(L)$ are obtained by replacing $f(x)$ by its asymptotic expansion.
At second order, we have
\be
f(x) =  \frac{1}{\pi}-\frac{x}{4}+\frac{7}{4}\,\frac{\zeta(3)}{\pi^3}\,x^2 + \dots
\ee
and we obtain
\ba
\label{eq:strongcheck}
\lim_{L\to\infty}\frac{\Delta^{\rm strong}_{\rm AF}(\lambda, L)}{L} &=& 1+\frac{\sqrt{\lambda}}{\pi} f\left(\frac{\pi}{\sqrt\lambda}\right) = 
\frac{\sqrt{\lambda}}{\pi^2}+\frac{3}{4} + \frac{7}{4}\,\frac{\zeta(3)}{\pi^2}\frac{1}{\sqrt\lambda} + \dots = \nonumber \\
&=& \frac{2\sqrt{2}}{\pi}\, g + \frac{3}{4} + \frac{7}{8\pi^3 \sqrt{2}}\,\zeta(3)\,\frac{1}{g} + \dots.
\ea
Here, $\Delta_{\rm AF}^{\rm strong}(\lambda, L)$ stands for the expansion Eq.~(\ref{eq:strongaf}).
Later, we shall compare this prediction with the finite $L$ analysis of the Lieb-Wu equations.

\subsection{The light states}

At the bottom of the perturbative multiplet there are light states. These are light in the sense that their anomalous
dimension is small in the perturbative region. These states are highly non trivial in the BDS description. In the Heisenberg
language, up to a change of sign in the anomalous dimension,  they are states which can be built adding many excitation over the antiferromagnetic state respecting the constraint
of zero spin and cyclicity. Not very much is known about these states since they correspond to non trivial 
distributions of Bethe roots in the thermodynamical limit~\cite{Sutherland}. In some cases, the Bethe Ansatz equations
can be solved at $L\to \infty$. An example is the lightest state 
which is associated with a limiting double contour distribution\cite{Beisert:2003xu,Beisert:2003ea}. For brevity,
we shall denote this state as $|{\rm FS}\rangle$ since in the BMN limit it is dual to the so-called 
folded string solution~\cite{Frolov:2003xy}.

The state $|{\rm FS}\rangle$ can be studied at finite $L$ and its anomalous dimension can be loop expanded.
If we express the result in terms of the 't Hooft coupling $\lambda$ we  find
\be
\Delta_{\rm FS}(\lambda, L) = L+\sum_{\ell\ge 1} \frac{c_\ell(L)}{L^{2\ell-1}}\,\lambda^{\ell}.
\ee
The reason for the various explicit powers of $L$ is that all the coefficients $c_\ell(L)$ have finite limits
as $L\to\infty$. Hence, the state $|{\rm FS}\rangle$ admits the BMN limit
\be
L\to\infty,\qquad \mbox{fixed}\ \frac{\lambda}{L^2} = 8\pi^2\left(\frac{g}{L}\right)^2.
\ee
It is usual to introduce the coupling $\lambda' = \lambda/J^2$ where $J=L/2$ is the angular momentum of the folded string.
The coupling $\lambda'$ is fixed in the BMN limit.
The BMN limit of the anomalous dimension  is then
\be
\label{eq:BMN}
\lim_{L\to\infty}\,\frac{1}{L}\,\Delta_{\rm FS}\left(\lambda' L^2, L\right) = F(\lambda'),\qquad \lambda'\ \mbox{fixed}.
\ee
The function $F(\lambda')$ has been first computed in~\cite{Frolov:2003xy}. At small $\lambda'$, it reproduces the gauge theory perturbative expansion
\be
F(\lambda') = 1 + \frac{0.7120}{8}\lambda' + \dots.
\ee
At large $\lambda'$, $F(\lambda')\simeq \frac{1}{\sqrt{2}}\,(\lambda')^{1/4}$, the typical behavior expected from AdS/CFT duality~\cite{Arutyunov:2004vx}.

In this paper, we work at finite $L$ and cannot access the limit Eq.~(\ref{eq:BMN}). Instead we are studying the $|{\rm FS}\rangle$ state
at fixed $L$, expanding $\Delta_{\rm FS}$ at large $\lambda'$. This is the same procedure we followed for the AF state
and is the kind of investigation described in~\cite{Minahan:2006zi}. We remark that there can be important differences
between this and the BMN limits. At finite $L$ and $\lambda'$ there can be terms with ambiguous $\lambda', L\to\infty$ limit.
An example could be, for instance,
\be
\label{correction}
\frac{\lambda'/L}{1+\lambda'/L},
\ee
tending to 1 when $\lambda'\to\infty$ and to 0 when $L\to \infty$.

BMN scaling  appears to be a general feature of the light states and is quite effective
in the search for the corresponding dual string states. 
It is natural to look at  BMN scaling  as an infinite volume limit where the fixed ratio $\sqrt{\lambda'}\sim g/L$ 
is interpreted as a finite size scaling variable. This smooth thermodynamical limit on the lattice 
can be explored more explicitly by evaluating in the Heisenberg model the SU(2) invariant correlation function 
\be
G_k = \langle\,{\rm FS}\,|\ \sigma_i\cdot\sigma_{i+k}\ |\,{\rm FS}\rangle.
\ee
In Fig.~(\ref{fig:corr}) we show its behavior at various $L$. Details of the calculation are reported in App.~A.
The correlation function $G_k$ expressed in terms of the scaled position $k/L-1/2$ 
tends to a smooth curve at large $L$. Similar results can be obtained for the other low lying states in the perturbative
multiplet. For completeness, we also show in Fig.~(\ref{fig:spectra}) the spectra of one loop anomalous dimensions for all $S=0$ cyclic states of the Heisenberg model
with even $8\le L\le 16$.

\section{Analysis of the full spectrum}

In this Section, we begin the analysis of the information that can be derived in the framework of the  full Hubbard model. 
In order to understand the general features of the weak to strong coupling flow
we analyze the full spectrum at variable $g$ and $L=4, 6, 8$. We follow the direct approach of Minahan~\cite{Minahan:2006zi} already applied to the case
$L=4$ that we also review to fix the approach, extending it to somewhat larger values of $L$. Later we shall discuss a different approach
based on the numerical solution of the Lieb-Wu equations.

\subsection{Review of the $L=4$ case}

As explained in the very nice investigation~\cite{Minahan:2006zi}, in the $L=4$ case and after restricting to cyclic states with $S=0$, 
there are only 6 remaining states. The antiferromagnetic state is non degenerate.
Its perturbative expansion involves only even powers of $g$ as an exact discrete symmetry of the model
\be
\label{eq:fourexpansion}
\Delta = 4 + \sum_{\ell\ge 1} c_\ell\,g^{2\ell}.
\ee
The full spectrum can be easily evaluated numerically and leads to the weak-strong coupling flow shown in Fig.~(\ref{fig:1}).

As explained in~\cite{Minahan:2006zi}, the perturbative expansion of $\Delta$ can be recovered quite efficiently from the 
expansion of the secular determinant
\be
P(\Delta, g) = \det\left(H_0+\frac{g}{\sqrt 2}\,H_1-\Delta\right).
\ee
One finds, 
\ba
P(\Delta, g) &=& \Delta^6-17\,\Delta^5+(119-16 g^2)\,\Delta^4+(-439+176 g^2)\,\Delta^3+(900-716g^2+32g^4)\Delta^2+ \nonumber \\
&+& (-972+1276 g^2-160 g^4)\Delta +432-840g^2+200g^4.
\ea
Replacing the expansion (\ref{eq:fourexpansion}) and matching the coefficients we find immediately
\ba
\Delta &=& 4 + 6\ g^2 - 12\ g^4 + 42\ g^6 - 318\ g^8 + 4524\ g^{10} - 
    63786\ g^{12} + 783924\ g^{14} - 8728086\ g^{16} + \nonumber \\
&+& 93893622\ g^{18} -     1038217494\ g^{20} + 12181236666\ g^{22} + \cdots
\ea

At strong coupling (large $g$) the leading behavior of the eigenvalues is linear in $g$ with a slope 
given by the eigenvalues of $H_1/\sqrt 2$. These are not immediately obtained from the dispersion relation
because not all multifermion states are allowed by the cyclic and $S=0$ conditions.
The explicit eigenvalues of the $6\times 6$ matrix $H_1/\sqrt 2$ can be computed analytically and are 
\be
0, 0, -2\,{\sqrt{2 - {\sqrt{2}}}}, 2\, {\sqrt{2 - {\sqrt{2}}}}, -2\, {\sqrt{2 + {\sqrt{2}}}}, 2\, {\sqrt{2 + {\sqrt{2}}}}.
\ee
From the free fermion dispersion relation Eq.~(\ref{eq:dispersion}) we see that the highest state has a slope 
that unambiguously identifies it with the free fermion state with the following level occupation
\be
\mathbf{n}_\uparrow = (0, 3),  \quad
\mathbf{n}_\downarrow = (0, 3).
\ee
The notation means that the components of $\mathbf{n}_\sigma$ are the energy modes of the $\sigma$ type fermions
according to Eq.~(\ref{eq:dispersion}).
This is nothing but the ground state of the Hubbard hopping term.

\subsection{Extension to longer operators}

The extension of the above direct approach to longer operators is in principle straightforward. However, some technical issues must be 
clarified in order to make the procedure systematic. We now illustrate the general features and then discuss the $L=6$ and $L=8$ cases.

To build the relevant states we first enforce the $S=0$ condition. Since $N_\uparrow = N_\downarrow$, we have automatically 
$S_z=0$. The operators $S^\pm$ are ineffective on doubly occupied sites. Also, they cannot move around the unpaired fermions.
As a consequence, we can partition the problem of listing $S=0$ states according to several sectors where we fix (a) 
the number and positions of the paired fermions, (b) the positions of the unpaired fermions.

Given such a sector, the fermions are no more itinerant from the point of view of the spin calculation. Then, spin zero states are 
simply the spin zero component of the SU(2) decomposition of the product of $N$ fundamental representations.
The actual wave function of these states is obtained by taking independent antisymmetrizations
with respect to pairs of up and down fermions. The independent antisymmetrizations are explicitly given by the 
SU(2) Young tableaux with two rows and $N/2$ columns. The Young tableaux entries in each column determine the independent
antysymmetrizations.

To give an example. Suppose that we have $3+3$ fermions on a $L=6$ lattice. In a sector where there is one paired couple 
in the rightmost site and the unpaired fermions are in the 1, 2, 3, 4 position, we have 6 states
\be
|*, *, *, *, 0, \uparrow\downarrow\rangle,\qquad * \equiv \uparrow\ \mbox{or}\ \downarrow,
\ee
with 3+3 fermions in total. Then, the relevant two Young Tableaux with their associated antisymmetrization prescriptions are
\be
\young(12,34)\qquad\young(13,24),
\ee
giving two spin zero states, in this sector. 

The number of Young tableaux with $S=0$ constructed with $L/2$ up and $L/2$ down fermions
is the first coefficient in (as usual $(-n)!\equiv 0$ when $n\in\mathbb{N}$)
\be
\left[\frac 1 2\right]^{\otimes 2N} = \bigoplus_{s=0, 1, 2, \dots} \frac{(2s+1)(2N)!}{(N+s+1)!(N-s)!}\,[s],
\ee
where $[s]$ is the spin $s$ representation of SU(2). Hence the desired number is the Catalan number $C_{L/2}$
\be
\left[\frac{1}{2}\right]^{\otimes L} = C_{L/2}\,\mathbf{0} \oplus\cdots, \quad C_{L/2} = \frac{L!}{(L/2)!(L/2+1)!}.
\ee
Summing over sectors with $p$ doubly occupied sites, we find the total number of $S=0$ states
\be
N_{S=0} = \sum_{p=0}^{L/2}\left(\begin{array}{c}L\\p\end{array}\right)
\left(\begin{array}{c}L-p\\L-2p\end{array}\right)\frac{(L-2p)!}{(L/2-p)!(L/2-p+1)!}
= \frac{L!(L+1)!}{((L/2)!(L/2+1)!)^2}.
\ee
This number is reduced roughly by the factor $1/L$ after the cyclic projection. This is implemented rather easily as follows.
We denote by $U$ the unitary operator which implements the shift symmetry. It can be checked that $U^L=1$ on half-filled states.
Cyclic states $|s\rangle$ satisfy $U|s\rangle = |s\rangle$. We then consider the shift symmetrizer
\be
S = \frac{1}{L}(1 + U + \cdots + U^{L-2} + U^{L-1}).
\ee
If ${\cal H}_0$ is the space of zero spin states and ${\cal C}_0$ the space of cyclic zero spin states, it is clear that the image
$S {\cal H}_0$
contains a basis of ${\cal C}_0$. Indeed, any state $|s\rangle\in{\cal C}_0$ can be written $|s\rangle = S|s\rangle$ and thus belongs to is in $S{\cal H}_0$.
Hence, we can compute the image $S {\cal H}_0$ and apply the Gram-Schmidt orthonormalization algorithm to simultaneously produce an orthonormal basis and 
also remove linearly dependent states.
Of course, this procedure can be applied to the states in each of the sectors that have been identified in the construction of 
$S=0$ states. This means, we repeat, sectors with fixed paired fermions and fixed positions of unpaired fermions. In addition, the cyclic structure of the 
fermion configurations in the cyclic states greatly helps in performing the orthonormalization by restricting to states with the same 
configurations modulo translations.

After these technical remarks, we analyze in turn the $L=6,8$ cases which, as we shall discuss, illustrate in our opinion some 
interesting features valid in more complicated cases.

\subsection{The non degenerate $L=6$ case}

After the spin zero and cyclic constraints, there is 1 state with no paired fermions.
Thus, the antiferromagnetic state is again non degenerate. Also, there are 10 states with two pairs, 14 states with three pairs
and 4 states with all fermions paired. The total dimension  is 29. 
Fig.~(\ref{fig:2}) illustrates the weak-strong coupling flow of the spectrum.

As compared with the $L=4$ cases, we observe several crossings of the coupling dependent levels. 
Such crossings are well known in integrable models, where they
are explained in terms of additional conserved charges commuting with the Hamiltonian~\cite{Levels}.

It is not feasible to evaluate the secular determinant $P(\Delta, g)$ at least if we do not want to resort to numerical 
evaluations. Instead, we can determine the analytical perturbative expansion of the highest 
eigenvalue by standard perturbation theory of non degenerate eigenvalues. 
Let $\psi_0$ be the normalized eigenvector of $H_0$ associated with the non degenerate eigenvalue $L$. 
Then, we set $\varepsilon_0 = L$ and 
iterate for $n\ge 1$
\ba
v_n &=& -H_1\, \psi_{n-1}+\sum_{k=1}^{n-1}\varepsilon_{n-k}\,\psi_k, \\
\varepsilon_n &=& -(\psi_0, v_n), \\
\psi_n &=& \frac{1}{H_0-L}\left(\varepsilon_n\,\psi_0+v_n\right).
\ea
The last equation is evaluated in the subspace $(\psi_0, \psi_n)=0$ where the (pseudo) inverse operator $(H_0-E_0)^{-1}$
exists. The perturbative expansion of the eigenvalue of $H_0 + \frac{g}{\sqrt 2} H_1$ is then
\be
\varepsilon(g) = \sum_{n\ge 0}\varepsilon_n \left(\frac{g}{\sqrt 2}\right)^n.
\ee
This algorithm is quite fast since it is based on matrix vector multiplications only. Applying it, we find 
\ba
\Delta &=& 6 + 6\,g^2 - 9\,g^4 + \frac{63\,g^6}{2} - \frac{621\,g^8}{4} + \frac{7047\,g^{10}}{8} - \frac{100953\,g^{12}}{16} + \nonumber \\ \nonumber \\ 
&+& \frac{2006127\,g^{14}}{32} - 
  \frac{46992069\,g^{16}}{64} + \frac{1100850183\,g^{18}}{128} - \frac{24465145473\,g^{20}}{256} + \\ \nonumber \\
&+& \frac{514257122079\,g^{22}}{512} -   \frac{10323764001117\,g^{24}}{1024} + \cdots \nonumber 
\ea
This expansion can be compared with the BDS approach based on the Heisenberg model~\cite{Beisert:2004hm}.
The agreement is perfect up to five loop order which is where the long-range Heisenberg Hamiltonian is  reliable
at $L=6$.
As we remarked, the Hubbard model calculation is conjectured to be exact at all orders in the loop expansions, although
a proof is lacking.
Of course, knowing the one-loop Bethe roots, the above expansion can also be obtained by perturbative expansion of the Lieb-Wu equations.
We do not insist on this point, since we are mainly concerned with strong coupling properties.

Again, we can exploit the direct diagonalization approach to identify the free fermion state to which 
the $g=0$ highest state flows. Comparing the slope of the highest eigenvalue (the maximum eigenvalue of $H_1/\sqrt 2$) 
with the dispersion relation Eq.~(\ref{eq:dispersion}) 
we find the two possibilities
\be
\mathbf{n}_\uparrow = (0, n, 5),  \qquad \mathbf{n}_\downarrow = (0, n', 5).
\ee
where $(n, n')$ can be $(1, 4)$ or $(4, 1)$. The contribution of these two fermions cancels in the energy.
Indeed, $\varepsilon_1+\varepsilon_4 = 0$. This means that we flow to an excited state of the full Hubbard model.
This is a clear consequence of the cyclic projection. Indeed, the ground state of the Hubbard hopping term for $L=6$ is not cyclic and is 
instead odd under the transformation Eq.~(\ref{eq:cyclic}).

The spin zero condition determines uniquely the correct combination of states which is the antisymmetric combination 
\be
\frac{1}{\sqrt{2}}(|0, 1, 5\rangle_{\uparrow}\otimes |0, 4, 5\rangle_\downarrow - |0, 4, 5\rangle_{\uparrow}\otimes |0, 1, 5\rangle_\downarrow )
\ee
As a check, we see that the largest eigenvalue of the explicit $29\times 29$ matrix $H_1$ on the 
cyclic spin zero states is non degenerate. 

\subsection{The degenerate $L=8$ case}

On a $L=8$ lattice there are 4900 half-filled states in the full Hubbard model. 
After the spin zero and cyclic constraints, there are 3 states with no paired fermions, 35 states with two pairs, 108 states with three pairs,
70 states with four pairs, and 10 fully paired states. The total dimension  is  thus reduced to 226. 
This is a remarkable reduction, but the dimension remains rather high. Nevertheless, we shall be able to complete the analysis.
The maximum eigenvalue of $H_0$ is threefold
degenerate. It contains the antiferromagnetic state and other two states with lower anomalous dimensions.
We postpone the discussion of flow. Again, 
it is clearly not feasible to evaluate the secular determinant. 
Also, we must deal with the complication that there are 3 states with eigenvalue $L=8$ at $g=0$.

We can determine the perturbative expansion of the highest 
eigenvalue by quantum mechanical formulae for perturbation theory of degenerate eigenvalues. In the case under consideration
the degeneration is removed at second order in $g$. A very simple practical algorithm is then the following.

Let $P_0$ be the projector onto the degenerate eigenspace ${\cal E}_0$ with eigenvalue $E_0=L=8$. Let 
\be
{\cal D} = P_0\,H_1\,(1-P_0)\frac{1}{H_0-E_0}(1-P_0)\,H_1\,P_0
\ee
be a $3\times 3$ matrix restricted to ${\cal E}_0$. Let its three eigenvectors be $\psi_0, \psi_0', \psi_0''$ in some order. They have 
distinct eigenvalues. We iterate for $n\ge 0$
\ba
\psi_{2n+1} &=& \frac{1}{H_0-E_0}\left(\sum_{k=1,\ {\rm odd}}^{2n-1}\varepsilon_{2n+1-k}\,\psi_k-H_1\,\psi_{2n}\right), \\
v_{2n+2} &=& \sum_{k=2, \ \rm even}^{2n}\varepsilon_{2n+2-k}\,\psi_k-H_1\,\psi_{2n+1}, \\
\varepsilon_{2n+2} &=& -(\psi_0, v_{2n+2}), \\
\psi_{2n+2} &=&  \frac{1}{H_0-E_0}\left(\varepsilon_{2n+2}\,\psi_0 + v_{2n+2}\right) + \alpha_{2n+2}'\,\psi_0' + \alpha_{2n+2}''\,\psi_0'' .
\ea 
The coefficients $\alpha_{2n}'\,\alpha_{2n}''$ are fixed by the condition $P_0 \, v_{2n-2} = 0$.

The explicit matrix $\cal D$ is rather complicated. Its eigenvalues are the three roots of the equation 
(of course in agreement with the one loop calculation in~\cite{Beisert:2003tq})
\be
\lambda^3 +40\,\lambda^2+464\,\lambda+1600 = 0. 
\ee
Finding numerically the eigenvectors and applying the above algorithm with the three possible
choices for $\psi_0$, we immediately find 
the three perturbative expansions
\ba
\Delta &=& 8 + 11.3022\,g^2 - 22.1706\,g^4 + 79.5035\,g^6 - 352.94\,g^8 + \\
&+& 1777.24\,g^{10} - 9743.47\,g^{12} + 56739.6\,g^{14} -825617\,g^{16}+ \cdots,  \nonumber \\
\Delta' &=& 8 + 5.45222\,g^2 - 7.94042\,g^4 + 31.2193\,g^6 - 159.093\,g^8 + \\
&+& 896.064\,g^{10} - 5378.33\,g^{12} + 33796.9\,g^{14} -222137\,g^{16}+ \cdots, \nonumber \\
\Delta'' &=& 8 + 3.24559\,g^2 - 1.88899\,g^4 + 1.2772\,g^6 + 1.0331\,g^8 + \\
&-& 8.2996\,g^{10} + 27.3006\,g^{12} - 70.9279\,g^{14} +159.235\,g^{16}+ \cdots \nonumber 
\ea
They are unavoidably numeric since they involve the algebraic irrational $\lambda$.

Again, we can compare with the BDS prediction at five loop. It is given in~\cite{Beisert:2004hm} terms of an 
algebraic number (denoted $\psi$)  which agrees with the above three $\lambda$ roots. The agreement is complete. Of course, beyond five-loop, 
the above expansions derived in the Hubbard model are new and must be checked against gauge theory perturbation theory.

Coming to the analysis of the weak-strong coupling spectrum flow, we see that it is quite complicated as illustrated in Fig.~(\ref{fig:3}).
The identification of the asymptotic free fermion state is less easy but straightforward. 
Let us denote by 
\be
|\Delta\rangle,\ |\Delta'\rangle,\ |\Delta''\rangle, 
\ee
the three states with eigenvalues expressed by the above expansion.
The highest state $|\Delta\rangle$ is the antiferromagnetic state and does not cross the other states along the flow. Instead, the other two degenerate states 
undergo several crossing as $g$ is increased.
However, it is easy to follow them along the crosses. We enumerate states starting from the highest. The two subleading states 
turns out to be 3rd and 8th asymptotic free fermion states. This is more clearly illustrated in Fig.~(\ref{fig:4})
where we show the first 9 ordered eigenvalues. Several crossings can be observed and in the end, the asymptotic eigenvalues
remain separated. Remarkably, at large $g$ the states $|\Delta'\rangle$ and $|\Delta''\rangle$
are close (in energy) to partner states  $|\widetilde\Delta'\rangle$ and $|\widetilde\Delta''\rangle$ that we shall now discuss.

As a first step toward the identification of the asymptotic states with free fermion states, we analyze the eigenvalues of 
the matrix $H_1/\sqrt{2}$ and compare them with the dispersion relation. If we denote by $s, s', s''$ the asymptotic 
slopes of the energies of the three states 
$|\Delta\rangle$,\ $|\Delta'\rangle$,\ $|\Delta''\rangle$, 
with respect to the coupling $g$, we find the following (unique) match
in terms of the free fermion energies Eq.~(\ref{eq:dispersion}), 
\ba
s &=& 2(\varepsilon_0+\varepsilon_1+\varepsilon_6+\varepsilon_7), \\
s' &=& 2(\varepsilon_0+\varepsilon_7), \\
s'' &=& 2(\varepsilon_1+\varepsilon_7).
\ea
The first relation permits to conclude that the highest states is nothing but the Hubbard model hopping term ground state. 
The other two relations identify two levels which are occupied by an up-down doublet $\uparrow\downarrow$ in momentum space.
The remaining 4 fermions (two up and two down) must be placed in the remaining levels with a total zero additional energy 
and respecting the cyclicity and zero spin conditions. This can only be achieved by leaving the four fermions unpaired and 
placing them symmetrical around the zero energy value. The level population is shown in Fig.~(\ref{fig:levels8}).

The first state on the left is the Hubbard model hopping ground state, as discussed. In the other states we have shown the two levels which are
unambiguously filled with a pair $\uparrow\downarrow$. We have also shown a particular admissible distribution of the remaining two up, and two down fermions 
on the allowed four symmetrical levels. We have circled them with dashed ellipses to emphasize that this is just one component of the exact state.
Indeed there are several possible distributions of the unpaired fermions. To further analyze, we take into account the spin zero condition. 
For both $|\Delta'\rangle$ and $|\Delta''\rangle$ the allowed states reduced to the two independent states
that are obtained by antisymmetrizing two pairs of up and down fermions.

Of course, there are two states because of the 
SU(2) decomposition
\be
{\bf\frac{1}{2}}\otimes{\bf\frac{1}{2}}\otimes{\bf\frac{1}{2}}\otimes{\bf\frac{1}{2}} = {\bf 2} \oplus {\bf 1}\oplus {\bf 1}\oplus {\bf 1}\oplus {\bf 0}\oplus {\bf 0}.
\ee
To be explicit, in the case of $|\Delta'\rangle$, two orthonormal states can be taken to be 
\ba
|e_1'\rangle &=& \frac{1}{2}(|\updownarrow,\updownarrow,\downarrow,\uparrow,\downarrow,\uparrow,0,0\rangle_p - |\updownarrow,\updownarrow,\downarrow,\uparrow,\uparrow,\downarrow,0,0\rangle_p + \\
&& - |\updownarrow,\updownarrow,\uparrow,\downarrow,\downarrow,\uparrow,0,0\rangle_p + |\updownarrow,\updownarrow,\uparrow,\downarrow,\uparrow,\downarrow,0,0\rangle_p), \nonumber \\
|e_2'\rangle &=& \frac{1}{2\sqrt{3}} (2|\updownarrow,\updownarrow,\downarrow,\downarrow,\uparrow,\uparrow,0,0\rangle_p - |\updownarrow,\updownarrow,\downarrow,\uparrow,\downarrow,\uparrow,0,0\rangle_p 
- |\updownarrow,\updownarrow,\downarrow,\uparrow,\uparrow,\downarrow,0,0\rangle_p + \nonumber \\
&& - |\updownarrow,\updownarrow,\uparrow,\downarrow,\downarrow,\uparrow,0,0\rangle_p - |\updownarrow,\updownarrow,\uparrow,\downarrow,\uparrow,\downarrow,0,0\rangle_p + 
2|\updownarrow,\updownarrow,\uparrow,\uparrow,\downarrow,\downarrow,0,0\rangle_p)
\ea
where the states $|\cdots\rangle_p$ are labeled with the fermion occupancy in momentum space and the momentum sites are ordered from the 
largest $\varepsilon_n$ ($\varepsilon_0$) to the smallest ($\varepsilon_4 = -\varepsilon_0$). 
With this notation, the Hubbard model hopping term ground state is $|\updownarrow, \updownarrow, \updownarrow, \updownarrow, 0, 0, 0, 0\rangle$.

Introducing analogous orthonormal states $|e''_{1,2}\rangle$ for the sector spanned by $|\Delta''\rangle$ and $|\widetilde\Delta''\rangle$,
the free fermion asymptotic states associated with $|\Delta'\rangle$ and $|\Delta''\rangle$ are suitable linear combinations 
of $|e'_{1,2}\rangle$ and $|e''_{1,2}\rangle$ that can be determined by perturbation theory in $H_0$.
In both cases, the relevant asymptotic state is the one with (slightly) larger energy as can be seen from Fig.~(\ref{fig:levels8}).

At first order, the degeneration is not removed. We find the same constant shift in both doublets. At second order, we find a non trivial energy splitting. 
We do not report the expression of the asymptotic eigenvectors which is really not useful. Instead, we give a closed form 
for the split eigenvalues.

If we denote by $\Delta'_\pm$ and $\Delta''_\pm$ the strong coupling expansion at second order of the doublets eigenvalues, we find 
\be
\Delta'_\pm = 2\,g\,(\varepsilon_0+\varepsilon_7)+\frac{23}{4}+\frac{1}{g}\,\delta'_\pm + \dots
\ee
where
\ba
\delta'_\pm &=& \frac{1}{128\sqrt{2}}\sqrt{882+584\sqrt 2+\sqrt{25906+16303\sqrt{2}}} \\
&\pm& \frac{1}{64}\sqrt{132+88\sqrt 2-2\sqrt{7330+5183\sqrt 2}} = \left\{\begin{array}{c}
0.301715 \\ 0.183563\end{array}\right.
\ea
The expansion of the second eigenvalue is instead
\be
\Delta''_\pm = 2\,g\,(\varepsilon_1+\varepsilon_7)+\frac{23}{4}+\frac{1}{g}\,\delta''_\pm + \dots
\ee
where
\ba
\delta''_\pm &=& \frac{1}{128}\sqrt{1026+79\sqrt 2+\sqrt{724177-\frac{157633}{\sqrt{2}}}} \\
&\pm& \frac{1}{16}\sqrt{10-3\sqrt 2-\sqrt{29-\frac{1}{\sqrt 2}}} = \left\{\begin{array}{c}
0.383745 \\ 0.300994\end{array}\right.
\ea
Notice that this is correct for $g>0$. Indeed, in general, the above strong coupling expansions must be written with $g\to |g|$
to respect the exact $g\to -g$ symmetry of the spectrum.

As a final comment, we remark that the above complicated expressions have been checked explicitly against the numerical
evaluation of the levels with full agreement.

\subsection{Summary of the results for $L=8$}

In conclusion, taking the upper states and evaluating the energy levels, we find the following result in the case $L=8$.
The AF state has been already discussed. The other two states in the perturbative multiplet have the following 
strong coupling expression of the anomalous dimensions
\ba
\Delta'_+ &=&  \sqrt{2\left(4+2\sqrt{2}+\sqrt{20+14\sqrt{2}}\right)}\ \ g+\frac{23}{4}+\frac{1}{g}\,\delta'_+ \dots, \\
\Delta''_+ &\equiv & \Delta_{\rm FS} = 2\sqrt{2+\sqrt{2+\sqrt{2}}}\ \ g+\frac{23}{4}+\frac{1}{g}\,\delta''_+ + \dots
\ea

\section{Direct analysis of the Lieb-Wu equations}

{\em A posteriori}, we can make some general comments on the previous analysis based on  direct diagonalization. We are considering an Hamiltonian
of the form 
\be
H = H_A + g\,H_B,
\ee
where $g$ is a coupling. As $g$ flows from 0 to $\infty$, each eigenstate of $H$ flows from an eigenstate of $H_A$ to an eigenstate of $H_B$.
It is clear that exact diagonalization permits to follow the flow for generic $H_A$, $H_B$, however the method is unrealistic for 
large dimension of the Hilbert space. The problem is very general and, as such, has no simple solution. Of course, 
what saves the day in our context is integrability. In principle, the Lieb-Wu equations can be solved for a particular state, {\em i.e.}
typical configuration of Bethe roots, without the need for huge calculations of eigensystems. Extending the calculation from $g=0$,
where the Bethe roots are those of the Heisenberg model, up to large $g$ should permit in principle to determine the strong coupling
behavior of any state.

However, it is also clear that the general task of solving the Lieb-Wu equations for all states in the perturbative multiplet and fixed $L$ (possibly large)
is not easy~\cite{BASolutions}. However, there are exceptions. These are the states where some general knowledge is available about the 
limiting distribution of Bethe roots at large $L$. In the next Sections, we shall discuss two important examples. The first is the AF state.
At half-filling, it is the unique state with completely real solutions to the Lieb-Wu equations. The second example is the lightest state $|{\rm FS}\rangle$.
Here, we know that at large $L$ the Bethe roots condense on two symmetric curves in the complex plane and we can exploit this information 
to evaluate them at least numerically. This case is considerably more difficult than the AF state because the Bethe Ansatz solution is complex.

Notice that in principle, one could use the original all-loop BDS equations. This calculation would be  reliable in the thermodynamical limit
including finite size corrections if needed. In this paper, we are concerned with finite $L$ properties, and therefore 
we have explored the explicit (numerical) solution of the more difficult Lieb-Wu equations.

\subsection{Real solution: the antiferromagnetic state}

At half filling, the AF state is described by the only genuine real solution of the Lieb-Wu equations. They read
\ba
L\,q_n &=& 2\pi\,I_n + 2 \sum_{j=1}^{L/2} \tan^{-1}\left[2(u_j-\sqrt{2}\,g\,\sin(q_n+\phi))\right], \\
2\pi\,J_k &=&  2 \sum_{j=1}^{L/2} \tan^{-1}(u_k-u_j) -2\sum_{m=1}^L \tan^{-1}\left[2(u_k-\sqrt{2}\, g\,\sin(q_m+\phi))\right],
\ea
where, in our problem, $n = 1,\dots, L$, and $k=1, \dots, L/2$.
We focus on the case $N = 4p$ where $\phi = \pi/(2L)$ and the Bethe quantum numbers are 
\ba
\{I_n\} &=& \{0, 1, 2, \dots, L-1\}, \\
\{J_k\} &=& \left\{-\frac{2p-1}{2}, -\frac{2p-3}{2}, \dots, \frac{2p-3}{2}, \frac{2p-1}{2}\right\}.
\ea
The iterative solution of the above equations is quite stable, as it is common when dealing with real solutions.
Following the evolution at $g\to\infty$ of the energy of the highest state, we have checked that it flows at strong coupling
to the ground state $|\psi_0\rangle$ of the Hubbard model hopping term. In momentum space, this is the state where all 
positive energy levels are doubly occupied
\be
|\psi_0\rangle = \prod_{n=1}^{L-1} \prod_{\sigma = \uparrow, \downarrow}\,a^\dagger_{\sigma, p_n} |0\rangle.
\ee
The AF state remains non degenerate at all couplings and we can apply first order perturbation theory to determine the 
first subleading correction at large $g$. Also, from the numerical solution of the 
Lieb-Wu equations, we can evaluate the finite $L$ next-to-subleading correction $\sim 1/g$.

Summing up, our result for the expansion of the anomalous dimension of the AF state at finite $L$ and large $g$ reads 
\be
\frac{\Delta_{\rm AF}(g, L)}{L} = \frac{\sqrt{2}}{L\,\sin\frac{\pi}{2L}}\, g + \frac{3}{4} + \delta_L\,\frac{1}{g} + \cdots
\ee
The first term is the energy of the Hubbard model hopping term ground state with twist. Its explicit formula comes from the sum
\be
2\sqrt{2}\sum_{n=-L/4}^{L/4-1}\cos\left(\frac{2\pi\,n}{L}+\frac{\pi}{2L}\right) = \frac{\sqrt{2}}{\sin\frac{\pi}{2L}}.
\ee
The coupling independent term is a universal constant. It is evaluated by computing the matrix element
\be
L-\frac{1}{L}\langle\psi_0| \sum_{p,q} a_{\uparrow, p}^\dagger\,a_{\uparrow, p}\,a_{\downarrow, q}^\dagger\,a_{\downarrow, q}| \psi_0\rangle
 = L -\frac{1}{L}\left(\frac{L}{2}\right)^2 = \frac{3}{4}\,L.
\ee
The next correction takes the numerical values reported in 
Tab.~(\ref{tab:1}).
\TABLE{
\begin{tabular}{|| cccccccc || }
\hline
$L$          & 8 & 12 & 16 & 20 & 24 & 28 & 32\\
$\delta_L$   & 0.0250979 & 0.0245631 & 0.0243433 & 0.0242308 & 0.0241652 & 0.0241234 & 0.024095 \\
\hline
\end{tabular}

\caption{Coefficient of the second order correction to the energy of the AF state at finite $L\in 4\mathbb{N}$ and large $g$.
A simple polynomial extrapolation at $L\to\infty$ gives $\lim_{L\to\infty} \delta_L = 0.0240(1)$.
}

\label{tab:1}
}
The above result is an exact expansion in inverse powers of $g$ at fixed $L$. As a non trivial check, it can be compared with
the result Eq.~(\ref{eq:strongcheck}) by taking the $L\to\infty$ limit term by term. We obtain
\be
\lim_{L\to\infty}\frac{\Delta^{\rm strong}_{\rm AF}(g, L)}{L} = \frac{2\sqrt{2}}{\pi}\, g + \frac{3}{4} + 0.0240(1)\,\frac{1}{g} + \dots.
\ee
This is in full agreement with Eq.~(\ref{eq:strongcheck}) since
\be
\delta_\infty\equiv \lim_{L\to \infty}\delta_L = \frac{7}{8\pi^3 \sqrt{2}}\,\zeta(3) = 0.0239866.
\ee

\subsection{A complex solution: the $|{\rm FS}\rangle$ state}

The general geometry and symmetry of the Bethe roots for the $|{\rm FS}\rangle$ state at large $L$ are of
great utility in determining them at finite $L\in 4\mathbb{N}$. As we shall discuss, there are 
numerical difficulties when $L=4p$ with even $p$, {\em i.e.} $L\,\mbox{mod}\,8 = 0$.
Instead the case $L=4p$ with odd $p$, {\em i.e.} $L\,\mbox{mod}\,8 = 4$, can be treated successfully.
In the next Sections we shall present our detailed results for the non trivial 
cases $L=12$, $20$.

\subsubsection{$L=12$}

At $L=12$ there are 14 states in the perturbative multiplet and many more extra states in the 
full Hubbard model.
We have determined the one loop Bethe roots for the $|{\rm FS}\rangle$ state as we now discuss. The six Bethe roots 
are non zero and symmetric under $u\to -u$. This reduces the problem to solving three polynomial equations in three variables.
Four roots are expected to be complex and the other two real.
Applying the resultant technique we find that the complex roots are among the solutions
of the following polynomial
\ba
R(x) &=& 1 - 291816\,x^2 + 2476695728\,x^4 - 4740875459840\,x^6 + 2716015001869568\,x^8 \nonumber \\
&& - 587934012140484608\,x^{10} + 53336517749102178304\,x^{12} - 
  1893188143985026269184\,x^{14} \nonumber\\
&& + 27036104777708125093888\,x^{16} - 119124909860572824600576\,x^{18} \nonumber\\
&& - 345397582972412910108672\,x^{20} + 
  1499936486421645590790144\,x^{22}\nonumber\\
&&  + 8744914427777415217414144\,x^{24} - 1155658862325646445510656\,x^{26} \\
&&  + 7466630646963993812402176\,x^{28} - 
  273168717774897644721668096\,x^{30}\nonumber\\
&&  + 46033164223912000241008640\,x^{32} - 219403347416043302531629056\,x^{34}\nonumber\\
&&  + 965423684859142279840923648\,x^{36} + 
  4267809644452908595622707200\,x^{38}\nonumber\\
&&  + 886394409006105612647399424\,x^{40} - 2540126101042720433221140480\,x^{42}\nonumber\\
&&  - 10295374756958736948626718720\,x^{44} - 
  27408343376808874282372300800\,x^{46}\nonumber\\
&&  - 16677333147947189938421760000\,x^{48} + 23395667776149943429890048000\,x^{50}\nonumber\\
&&  + 48702367142746624075235328000\,x^{52} + 
  43027672720198395903344640000\,x^{54}\nonumber\\
&&  + 21488277238759969133690880000\,x^{56} + 5377757322242364014592000000\,x^{58}\nonumber\\
&&  + 496408368206987447500800000\,x^{60}\nonumber.
\ea
This polynomial contains several spurious solutions and is perhaps not the most economical choice.
Nevertheless, it contains the exact roots and we quote it for the reader's convenience.~\footnote{
A single polynomial for the case $L=8$ is quoted in~\cite{Beisert:2004hm}. However, there is a misprint
in one coefficient. The correct resolvent is $
R(x) = -1 + 648\ x^2 - 36464\ x^4 + 81664\ x^6 - 16128\ x^8 + 460800\ x^{10} + 552960\ x^{12}.$
}
The roots associated with the $|{\rm FS}\rangle$ state are 
\ba
u_1 &=& -u_2 = \alpha, \\
u_3 &=& -u_4 = \overline\alpha, \\
u_5 &=& -u_6 = \beta, 
\ea
where
\ba
\alpha &=& 0.6762450414055523 + 0.9936333912043784\ \ i, \\
\beta &=& 0.6780174422473694,
\ea
in agreement with the results in~\cite{Beisert:2003xu}. We
start from this solution plus the condition for the momenta $q$
\be
q_n = \frac{2\pi}{L}(n-1),\quad 1\le n\le L,
\ee
which is valid at $g=0$. Then, we increase $g$ and determine step by step the new solution of the full Lieb-Wu equations. 
This procedure is numerically stable and permits to determine the energy flow as well as the change in the Bethe momenta and $u$ variables.

We show in Fig.~(\ref{fig:bethe12u}), the evolution of the Bethe Ansatz solution $\{u_i\}$ as $g$ is increased up to $g=15$. More interestingly, 
we show in Fig.~(\ref{fig:bethe12q}), the evolution of the momenta. The flow permits to derive the asymptotic occupancy in the 
free fermion limit. In terms of the indices $n$, the final occupation of states is as follows. There are singly occupied modes with mode
indices
\be
n = 0, 11, 2, 9, 3, 8, 5, 6,
\ee
and two doubly occupied levels at modes
\be
n = 1, 10.
\ee
As in the $L=8$ case, the singly occupied levels contain 4 up fermions and 4 down fermions in a $S=0$ combination. There are
several possibilities and only one can be selected by perturbation theory. We do not pursue the strong coupling
correction analytically. Instead we determine the leading term at large $g$ and the subleading contribution that can be obtained by 
first order perturbation theory. The leading contribution (the coefficient of $g$ at large $g$) is 
\be
 2(\varepsilon_1 + \varepsilon_{10}) = 2\sqrt{2}\left(\cos\frac{5\pi}{24}+\cos\frac{41\pi}{24}\right) = 
4\,\cos\frac{\pi}{24} = 2\sqrt{2+\sqrt{2+\sqrt{3}}}.
\ee
The subleading term can be computed analytically and is $26/3$. Hence we have found that for the $|{\rm FS}\rangle$ state at $L=12$
we have 
\be
\Delta_{\rm FS} = 2\sqrt{2+\sqrt{2+\sqrt{3}}}\,\,g + \frac{26}{3} + 0.597(1)\,\frac{1}{g}+\cdots,
\ee
where we have also indicated the fitted coefficient of the NNLO term in the strong coupling expansion.
The agreement with the calculated energy is shown in Fig.~(\ref{fig:bethe12e}).

The asymptotic fermion configuration is shown in Fig.~(\ref{fig:levels12}) where we simply draw the single particle level
and their occupation without specifying the spin of the singly occupied levels.

\subsubsection{$L=16$}

We can repeat the analysis for $L=16$. In this case, we failed to obtain an exact resultant encapsulating the exact 
one loop Bethe roots. Instead, we have computed them numerically. The symmetry of the 8 roots is 
\ba
u_1 &=& -u_2 = \alpha, \\
u_3 &=& -u_4 = \overline\alpha, \\
u_5 &=& -u_6 = \beta,  \\
u_7 &=& -u_8 = \overline\beta, 
\ea
where
\ba
\alpha &=&  0.9011983985707239 + 0.5000879064837407\,i, \\
\beta  &=&  0.915478863907937  + 1.4850185722704357\,i.
\ea
The imaginary part of $\alpha$ is quite near to $\frac{1}{2}$. This is a source of instability in the solution of the Lieb-Wu equations.
Indeed, as $g$ is increased, we numerically observe that four of the Bethe roots tend quickly to a singular configuration.
This problem does not occur if $L\,\mbox{mod}\, 8 = 4$. We do not try to deal with the singularities of the $L=16$ case and 
instead study the more involved, but more stable case $L=20$.

\subsubsection{$L=20$}

The 10 one-loop Bethe roots satisfy the following conditions:
\ba
u_1 &=& -u_2 = \alpha\in\mathbb{R}_{>0}, \\
u_3 &=& -u_4 = \beta, \\
u_5 &=& -u_6 = \overline\beta, \\
u_7 &=& -u_8 = \gamma,  \\
u_9 &=& -u_{10} = \overline\gamma.
\ea
We find the following numerical solution
\ba
\alpha &=&  1.1309564538305164, \\
\beta  &=&  1.1310261843923932 + 0.9998455911389437\,i, \\
\gamma  &=& 1.1784184821892867 + 1.980402937535511\,i.
\ea
We show in Fig.~(\ref{fig:bethe20u}), the evolution of the Bethe Ansatz solution $\{u_i\}$ as $g$ is increased up to $g\simeq 18$. 
Fig.~(\ref{fig:bethe20q}) shows the evolution of the momenta. Again, we can derive the asymptotic occupancy in the 
free fermion limit. In terms of the indices $n$, the singly occupied modes are
\be
n = 0, 1, 3, 4, 5, 6, 7, 9, 10, 11, 13, 14, 15, 16, 18, 19,
\ee
and there are again 2 doubly occupied levels at modes
\be
n = 2, 17.
\ee
The singly occupied levels contain 8 up fermions and 8 down fermions in a $S=0$ combination. 
The leading contribution to the anomalous dimension (the coefficient of $g$ at large $g$) is 
\be
2(\varepsilon_2 + \varepsilon_{17}) = 2\sqrt{2}\left(\cos\frac{9\pi}{40}+\cos\frac{69\pi}{40}\right) = 4\cos\frac{\pi}{40}.
\ee
The subleading term can be computed analytically and is $73/5$. Hence, in summary, the $|{\rm FS}\rangle$ state at $L=20$
admits the strong coupling expansion
\be
\Delta_{\rm FS} = 4\,\cos\frac{\pi}{40}\,\,g + \frac{73}{5} + 0.953(1)\,\frac{1}{g}+\cdots,
\ee
where we have also indicated the fitted coefficient of the NNLO term in the strong coupling expansion.
As before, we shown the agreement with the calculated energy is shown in Fig.~(\ref{fig:bethe20e}).

The asymptotic fermion configuration is completely similar to the $L=12$ case. The doubly occupied levels are in the middle of
the single particle positive energy levels. The other positive energy levels are singly occupied, as well as their mirror 
levels with negative energy.

\subsubsection{Conjecture for the $|{\rm FS}\rangle$ state at general $L = 4(2k+1)$}

The results at $L=12$ and $20$ are quite symmetric and completely similar. It is natural to conjecture that for all $L=4(2k+1)$, 
the pattern is identical. This means that the $|{\rm FS}\rangle$ state is obtained at strong coupling as
the state with the following properties.
\begin{enumerate}
\item The positive energy single fermion levels are all occupied with one fermion, with the exception of the
central levels with mode numbers
\be
n = k, \ L-k-1.
\ee
These are doubly occupied.
\item The negative energy levels which are mirror of singly occupied levels are also singly occupied.
\item The negative energy levels which are mirror of doubly occupied levels are empty.
\end{enumerate}
Evaluating the leading and subleading contributions to the anomalous dimension gives the strong coupling 
expansion
\be
\label{eq:conj}
\Delta_{\rm FS}(g, L) = 4\,\cos\frac{\pi}{2L}\,g + \frac{3L^2-2L+8}{4L} + {\cal O}\left(\frac{1}{g}\right).
\ee
Eq.~(\ref{eq:conj}) is expected to be the exact strong coupling expansion at any fixed $L =4(2k+1)$.
We conclude with a comment about the other cases $L=4(2k)$. The explicit results at $L=8$ and preliminary data at $L=16$
suggest that the $|{\rm FS}\rangle$ state is again flowing to a state like the above with the same expressions for the leading and subleading
terms in the strong coupling expansion. Indeed, at $L=8$, the above parametrization reproduces the exact result that we derived by 
exact diagonalization. However, we have not enough empirical support to firmly establish this result.

\section{Summary of results and discussion}

To summarize, we report our main results for the large $g$ expansion at fixed $L$ of the anomalous dimension
of the antiferromagnetic and $|{\rm FS}\rangle$ states.
For the antiferromagnetic operator we have found
\be
\frac{\Delta_{\rm AF}(g, L)}{L} = \frac{\sqrt{2}}{L\,\sin\frac{\pi}{2L}}\, g + \frac{3}{4} + \delta_{{\rm AF}, L}\,\frac{1}{g} + \cdots.
\ee
The first two terms are exact. About the last one, we have  shown how to compute it for large $L$. We have also provided the asymptotic limit
\be
\delta_{{\rm AF}, \infty} = \frac{7}{8\pi^3 \sqrt{2}}\,\zeta(3).
\ee
For the folded string dual, we have found (at finite  $L\,\mbox{mod}\,8 = 4$)
\be
\label{xxx}
\Delta_{\rm FS}(g, L) = 4\,\cos\frac{\pi}{2L}\,g + \frac{3L^2-2L+8}{4L} + \delta_{\lambda, L}\,\frac{L}{g}+\cdots,
\ee
where we have explicitly computed
\be
\delta_{\lambda, 12} = 0.0498(1), \ \ 
\delta_{\lambda, 20} = 0.0477(1).
\ee
In particular, the leading and subleading terms of these expressions are exact at all finite $L$. As such they are beyond the region of applicability
of the BDS approximation which, due to wrapping terms, is limited to $L\to \infty$. Indeed, they are a genuine result provided by the Hubbard model framework
where they express properties of specific free fermion states.

As we remarked, our expansions are obtained by taking $\lambda'$ large at fixed $L$. 
An important issue if then the comparison with the BMN limit. Indeed, as pointed out by Minahan~\cite{Minahan:2006zi}, 
it is not totally clear how to recover the $(\lambda')^{1/4}$ behavior of string states from the strong coupling expansion
of the Hubbard model. The tricky proposal in~\cite{Minahan:2006zi} is based on the assumption that 
at large $g$ there are  doubly occupied levels with small single particle energy of order $1/L$. Unfortunately, we have seen
that this is not valid for the considered states with many excitations. The doubly occupied levels give an asymptotic slope $\Delta/g$ which is 
of order 1 for the minimal energy state $|{\rm FS}\rangle$.

The BMN limit is obtained by fixing $\lambda' = \lambda/J^2$ and taking $J\to \infty$ where $J=L/2$ is the angular momentum of the semiclassical dual state.
We have seen that at finite $L$ there can be correction terms (like Eq.~(\ref{correction})) with an ambiguity in the $L, \lambda'\to \infty$ limit. 
If we want to compare with the BMN limit we have to enforce the correct ordering and require $\lambda'\ll L$.
Let us see the role of this constraint in the case of our data at $L=12$ and $20$. The BMN anomalous dimension 
of the folded string can be written 
\be
\lim_{L\to \infty}\frac{\Delta_{\rm FS}(\lambda' L^2, L)}{L} \equiv F(\lambda') = \frac{1}{2} K(q)\left[\frac{4q\lambda'}{\pi^2} + \frac{1}{E(q)^2}\right]^{1/2},
\ee
where $q = q(\lambda')$ is the solution of 
\be
\frac{4\lambda'}{\pi^2} = \frac{1}{(K(q)-E(q))^2}-\frac{1}{E(q)^2},
\ee
and $K(q)$, $E(q)$ are standard complete elliptic integrals of the first and second kind.

We show in Fig.~(\ref{fig:bmn}) the comparison of $F(\lambda')$ and the ratio $\Delta/L$ for the lightest state at $L=12$ and $20$.
The left panel shows that there is good agreement for  $\lambda'$ up to about $3-4$, reasonably within the BMN scaling 
window. This is a rather large value suggesting that the agreement is working beyond perturbation theory. Indeed, we show in the same figure
the 8th and 9th order perturbative expansions of $F(\lambda')$ which read
\ba
\lefteqn{F(\lambda') = } && \\
&& 1 + 
    0.089004 \ \lambda' - 
    0.013272 \ (\lambda')^2 + 
    0.002839  \ (\lambda')^3 - 
    0.000676  \ (\lambda')^4 + 
    0.000172   \ (\lambda')^5 \nonumber \\
&& - 
    0.000047    \   (\lambda')^6 + 
    0.000014    \ (\lambda')^7 - 
    4.315511         \cdot 10^{-6}\ (\lambda')^8 + 
    1.424401            \cdot 10^{-6}\ (\lambda')^9 + \dots \nonumber.
\ea
The two curves suggest a convergence radius around $\lambda'\simeq 3$, somewhat smaller than the 
region of agreement. If so, this could be a signal that we are slowly recovering the BMN result. 

\section{Conclusions}

In this paper we have considered a particular class of gauge invariant operators in the
planar limit of  ${\cal N}=4$ SYM. These are length $L$ single trace operators in the SU(2) sector
with zero spin. Remarkable members of this class are the so-called antiferromagnetic operator and
the dual of the semiclassical folded string state.

We have assumed a recently proposed relation between the gauge theory and 
a Hubbard-like model of itinerant fermions. This approach permits to evaluate the anomalous dimensions
of the gauge invariant operators at all couplings. In particular, we access the strongly coupled region 
at fixed $L$.

Our investigation has been based on two complementary techniques. First, we have evaluated the full spectrum
of the model for operators with $L=4, 6, 8$. This has provided useful information about the 
mutual relation between the states of the Hubbard model and the perturbative multiplet of states 
with a clean relation to gauge invariant operators. 

Then, we have investigated the numerical solutions of the Lieb-Wu equations. They are a 
powerful tool that permits, in principle, to follow a particular state from weak to strong coupling in a 
totally controlled way.
Our results are very simple explicit formulas for the strong coupling expansion
of the anomalous dimensions. They are expressed in terms of specific free fermion states whose properties are easily
computable. 

We hope that this investigation will be useful to shed some additional light over the non perturbative 
features of the states in the multiplet as well as over the connection between the gauge theory and the 
underlying integrable Hubbard model.
This work can be extended in several directions. Some are rather obvious like (a) $S\neq 0$ states in the SU(2)
sector, (b) sectors different than SU(2), (c) other particular states like for instance the dual of the semiclassical 
circular string solution. In principle, it should be possible to study the strong coupling region by 
direct perturbative expansion of the Lieb-Wu equations although this is a delicate analysis~\cite{Arutyunov:2006av}.
This could be a valuable effort. It would be very nice to reproduce in some limit of  the Hubbard model
the true strong coupling behavior of string states, {\em i.e.} the typical relation $\Delta\sim(\lambda')^{1/4}$
for the light states. This is non trivial at the numerical level since large $\lambda'$ and irrelevance of
possible corrections like (\ref{correction}) would require quite large lattice sizes $L$.

The most interesting extension seems to be a detailed study of the other light states in the perturbative multiplet,
perhaps exploiting in a deeper way their {\em nearly-BPS} nature~\cite{Mateos:2003de}.
As $L$ increases, our investigation shows that there is a growing number of states with a smooth 
$L\to \infty$ limit. This is not unexpected and they should be described by a suitable effective theory in the continuum.
This line of analysis have been discussed in~\cite{Continuum} in the context of the loop-corrected Heisenberg model and should be 
extended to the Hubbard model~\cite{Roiban:2006jt}.
For these states, the finite $L$ analysis of their associated Lieb-Wu solution 
could provide some hindsight on the possible limiting distribution of Bethe roots and suggest a strategy to 
evaluate their thermodynamical limit. In the end, this could lead to new examples of AdS/CFT specific dualities.
Indeed, the search of new string states dual to novel Bethe Ansatz solutions seems to be far from the end~\cite{Roiban:2006jt}.

\medskip
\noindent
{\bf Acknowledgments}:
We thank G. F. De Angelis for conversations about rigorous results on lattice fermion models.

\appendix

\section{Exact diagonalization of the $S=0$ sector of the Heisenberg model}

In this Appendix, we report some useful techniques for exact diagonalization of Heisenberg-like models in 
sectors with fixed SU(2) spin. As discussed in the analysis of RSS, the two loop dilatation operator is
\be
{\cal D} = L + \frac{g^2}{2}\sum_i(1-\sigma_i\cdot\sigma_{i+1})-\frac{g^4}{4}\sum_i(3-4\sigma_i\cdot\sigma_{i+1}+\sigma_i\cdot\sigma_{i+2}) + \cdots,
\ee
where we assume periodic boundary conditions. This operator acts in the perturbative multiplets, i.e. on the $S=0$ cyclic states of the Heisenberg
spin model. We can write the various $\sigma_i\cdot\sigma_j$ terms by means of transpositions operators flipping spins at sites $i$, $j$
\be
P_{i,j} = \frac{1}{2}(1+\sigma_i\cdot\sigma_j).
\ee
The dilatation operator can be rewritten
\be
{\cal D} = L + g^2 \sum_i(1- P_{i, i+1}) + \frac{g^2}{2}\sum_i(4P_{i,i+1}-P_{i, i+2}-3) + \cdots
\ee
where we notice that (under periodic identification of the boundaries)
\be
P_{i,\,i+2}=P_{i,i+1}\,P_{i+1,\,i+2}\,P_{i,\,i+1},
\ee
and the full operator is written in terms of elementary transpositions only, i.e. transpositions of nearest neighboring spins. 
Also, cyclic states are states invariant under
a lattice shift $T$ that can also be written in terms of elementary transpositions as the product
\be
T = P_{L-1, L}\,\cdots\,P_{2,3}\,P_{1,2} .
\ee
We are interested in diagonalizing the dilatation operator in the $S=0$ sector. 
The states in this sector are associated with standard SU(2) Young tableaux with two rows and $L/2$ columns, as already discussed.
However, now we are no more interested  in the detailed spin positions and we do not need translating the YT in explicit (anti)symmetrized states. 
Instead, we can exploit an old computationally efficient  parametrization of orthogonal states~\cite{Symmetric} which turns out to be very suitable for our problem.

We associate a state $|{Y}\rangle$ to each distinct spin zero Young tableaux ${Y}$. 
Then, the nearest-neighbor transposition $P_{k, k+1}$ has the following matrix elements
\be
\langle {Y}' | P_{k, k+1} | {Y}\rangle = 
\left\{
\begin{array}{cc}
\rho_k({Y}) \equiv \displaystyle \frac{1}{d_k({Y})}, & {Y} = {Y}' \\ \\
\sqrt{1-\rho^2} , & {Y}'\  \mbox{is obtained from}\ {Y}\ \mbox{by}\  k\leftrightarrow k+1 \\ \\
0, & \mbox{otherwise}.
\end{array}
\right.
\ee
The number $d_k({Y})$ is the {\em axial distance} between the labels $k$ and $k+1$ inside $Y$. It is defined as the sum of steps
which are required to move from $k$ to $k+1$. The steps are positive on the right and upward and negative otherwise.

The above representation of states is not very convenient for the analysis of generic correlation functions.
However, it is quite efficient for energies and SU(2) invariant correlation functions.

\newpage
\vskip 2cm
\FIGURE{\epsfig{file=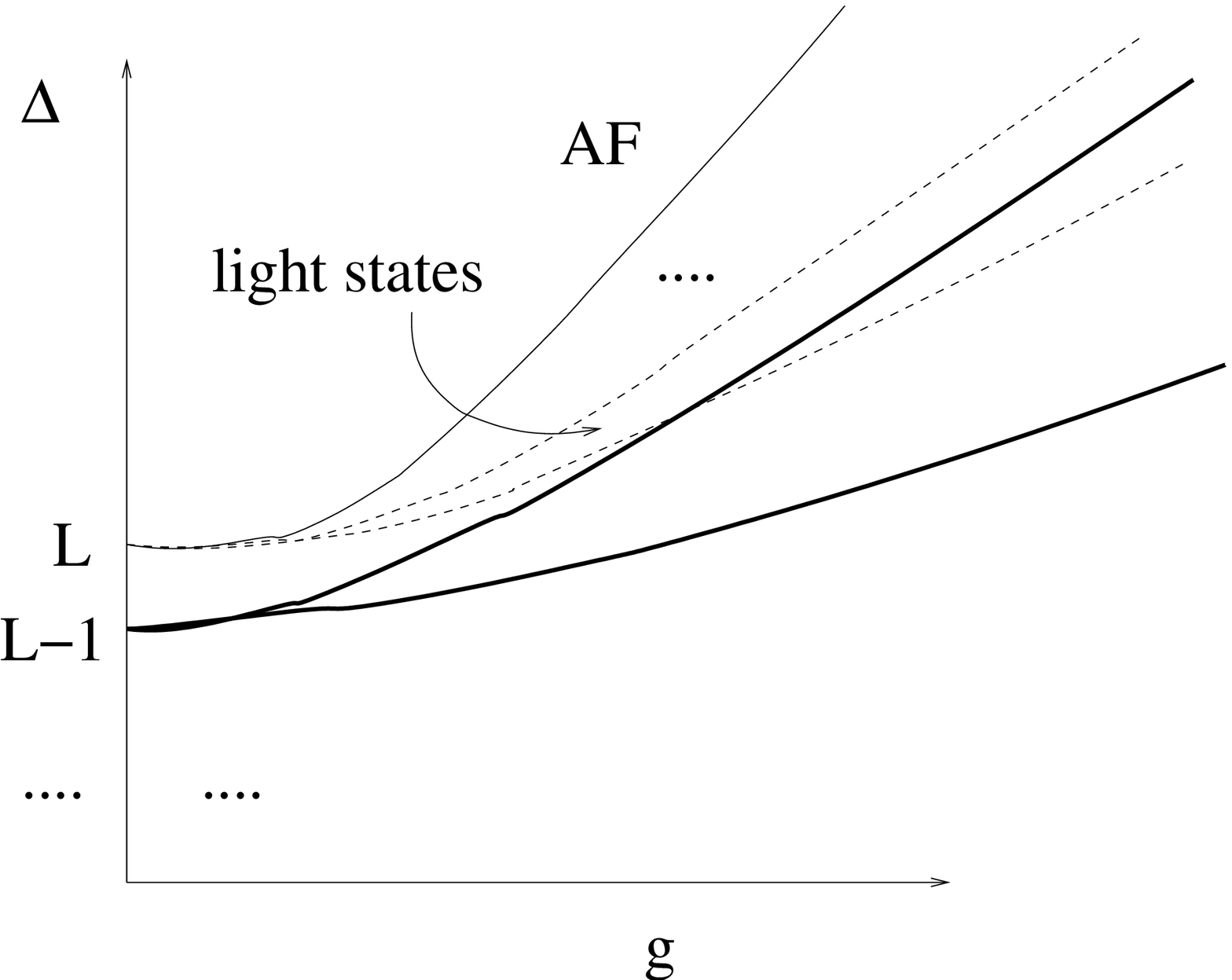,width=10cm} 
        \caption{Qualitative description of the Hubbard model spectrum. The upper line is the AF state. The lowest dashed
lines in the perturbative multiplet are the {\em light} states. We also show two states from a different multiplet. One of them
crosses a light state. At large $g$ all lines are linear in $g$. At small $g$, $\Delta(g) = L + {\cal O}(g^2)$ in the perturbative
multiplet.}
	\label{fig:qualitative}}

\newpage

\vskip 1.45cm
\FIGURE{\epsfig{file=Figures/HeisenbergCorr.eps,width=12cm} 
        \caption{Spin correlation function for the lightest $|{\rm FS}\rangle$ state of the Heisenberg chain (under the 
cyclic and $S=0$ constraints).}
	\label{fig:corr}}

\vskip 1.45cm
\FIGURE{\epsfig{file=Figures/HeisenbergSpectra.eps,width=12cm} 
        \caption{Spectra of one loop anomalous dimensions at even $8\le L \le 16$. The anomalous dimension of the n-th state
is written $\Delta_n = L + \Delta_n^{(1)}\,g^2 + \cdots$ and the plot shows $\Delta^{(1)}_n$.}
	\label{fig:spectra}}

\newpage
\vskip 2cm
\FIGURE{\epsfig{file=Figures/L4flow.eps,width=10cm} 
        \caption{Spectrum flow for the Hubbard model at $L=4$. The boldface line is the highest eigenvalue.}
	\label{fig:1}}

\vskip 1.45cm
\FIGURE{\epsfig{file=Figures/L6flow.eps,width=10cm} 
        \caption{Spectrum flow for the Hubbard model at $L=6$. The boldface line is the highest eigenvalue.}
	\label{fig:2}}

\newpage

\vskip 1.45cm
\FIGURE{\epsfig{file=Figures/L8flow.eps,width=10cm} 
        \caption{Spectrum flow for the Hubbard model at $L=8$.}
	\label{fig:3}}

\vskip 1.45cm
\FIGURE{\epsfig{file=Figures/detailedL8flow.eps,width=10cm} 
        \caption{Spectrum flow for the Hubbard model at $L=8$. A detailed view of the highest 9 eigenvalues.}
	\label{fig:4}}

\newpage

\vskip 1.45cm
\FIGURE{\epsfig{file=Figures/levels.eps,width=14cm} 
        \caption{Asymptotic free fermion states for the perturbative multiplet at $L=8$. 
As explained in the text, the fermions encircled by dashed ellipses indicate 
one of the various (spin) components of the actual state.}
	\label{fig:levels8}}

\newpage

\vskip 1.45cm
\FIGURE{\epsfig{file=Figures/LiebWu.L12.u.eps,width=12cm} 
        \caption{State $|{\rm FS}\rangle$ at $L=12$. Evolution of the Bethe parameters $u_1, \dots, u_6$ up to $g=15$.}
	\label{fig:bethe12u}}

\vskip 1.45cm
\FIGURE{\epsfig{file=Figures/LiebWu.L12.q.eps,width=12cm} 
        \caption{State $|{\rm FS}\rangle$ at $L=12$. Evolution of the Bethe parameters $q_1, \dots, q_{12}$ up to $g=15$.}
	\label{fig:bethe12q}}

\newpage

\vskip 1.45cm
\FIGURE{\epsfig{file=Figures/LiebWu.L12.energy.eps,width=12cm} 
        \caption{State $|{\rm FS}\rangle$ at $L=12$. Coupling dependence of $\Delta_{\rm FS}(g)$.}
	\label{fig:bethe12e}}

\vskip 1.45cm
\FIGURE{\epsfig{file=Figures/levels12.eps,width=12cm} 
        \caption{State $|{\rm FS}\rangle$ at $L=12$. Fermion configuration at strong coupling.}
	\label{fig:levels12}}

\newpage

\vskip 1.45cm
\FIGURE{\epsfig{file=Figures/LiebWu.L20.u.eps,width=12cm} 
        \caption{State $|{\rm FS}\rangle$ at $L=20$. Evolution of the Bethe parameters $u_1, \dots, u_{10}$ up to $g=18$.}
	\label{fig:bethe20u}}

\vskip 1.45cm
\FIGURE{\epsfig{file=Figures/LiebWu.L20.q.eps,width=12cm} 
        \caption{State $|{\rm FS}\rangle$ at $L=20$. Evolution of the Bethe parameters $q_1, \dots, q_{20}$ up to $g=18$.}
	\label{fig:bethe20q}}

\newpage

\vskip 1.45cm
\FIGURE{\epsfig{file=Figures/LiebWu.L20.energy.eps,width=12cm} 
        \caption{State $|{\rm FS}\rangle$ at $L=20$. Coupling dependence of $\Delta_{\rm FS}(g)$.}
	\label{fig:bethe20e}}

\vskip 1.45cm
\FIGURE{\epsfig{file=Figures/BMN.eps,width=12cm} 
        \caption{Ratio $\Delta/L$ for the state $|{\rm FS}\rangle$ as a function of $\lambda'$ at $L=12$, $20$, and in the BMN limit $L\to\infty$.}
	\label{fig:bmn}}

\end{document}